# Inverted Perovskite Photovoltaics using Flame Spray Pyrolysis Solution based CuAlO$_2$/Cu-O Hole Selective Contact

By Achilleas Savva,[1] Ioannis T. Papadas,[1] Dimitris Tsikritzis,[1] Apostolos Ioakeimidis,[1] Fedros Galatopoulos,[1] Konstantinos Kapnisis,[1] Roland Fuhrer,[2] Benjamin Hartmeier,[2] Marek F. Oszajca,[2] Norman A. Luechinger,[2] Stella Kennou,[3] Gerasimos Armatas[4] and Stelios A. Choulis*,[1]

[1] Dr A. Savva, Dr I.T. Papadas, Dr Dimitris Tsikritzis,  Mr Apostolos Ioakeimidis,  Mr Fedros Galatopoulos, Dr Konstantinos Kapnisis, Prof S. A. Choulis,

Molecular Electronics and Photonics Research Unit, Department of Mechanical Engineering and Materials Science and Engineering, Cyprus University of Technology, Limassol, 3603 (Cyprus).

[2] Roland Fuhrer, Benjamin Hartmeier, Marek Oszajca, Dr Norman Luechinger

Avantama Ltd, Staefa, Laubisrutistr. 50, CH-8712, Switzerland

[3] Department of Chemical Engineering, University of Patras, 26504, Patras, Greece

[4] Department of Materials Science and Technology, University of Crete, Heraklion 71003, Greece

*Corresponding Author: Prof. Stelios A. Choulis

E-mail: stelios.choulis@cut.ac.cy

**KEYWORDS:** CuAlO$_2$/Cu-O alloying, Delafossite oxides, hole transporting layers, perovskites solar cells, flame spray pyrolysis

**ABSTRACT:**

We present the functionalization process of a conductive and transparent $CuAlO_2$/Cu-O hole transporting layer (HTL). The $CuAlO_2$/Cu-O powders were developed by flame spray pyrolysis and their stabilized dispersions were treated by sonication and centrifugation methods. We show that when the supernatant part of the treated $CuAlO_2$/Cu-O dispersions is used for the development of $CuAlO_2$/Cu-O HTLs the corresponding inverted perovskite–based solar cells show improved functionality and power conversion efficiency of up to 16.3% with negligible hysteresis effect.

## 1. INTRODUCTION

The discovery of novel solution processed light-harvesting materials boosted the photovoltaic research towards innovative concepts, such as hybrid perovskite-based solar cells.[1] Within less than 10 years of development, perovskite-based solar cells exhibit lab-scale power conversion efficiencies (PCEs) of over 22%.[2,3,4] In addition, the substantial progress on novel perovskite solar cell architectures based on solution processing, marks the compatibility of these devices with large scale printing production.[5] Among many types of perovskite solar cell architectures,[6,7] the planar heterojunction p-i-n architecture show some relevant advantages to large scale production. For example, the fullerene processed on top of the perovskite as the n-contact has been found to eliminate the current-voltage (J-V) hysteresis, which is generally observed in the n-i-p device architecture.[8] Furthermore, both the fullerene n-type contact as well as the p-type contact can be fabricated at low processing temperatures.[9,10]

Charge selective contacts are of high importance for the functionality of p-i-n perovskite solar cells.[11,12] Poly(3,4-ethylenedioxythiophene)-poly(styrenesulfonate) (PEDOT:PSS) is the most commonly used p-type material acting as hole transporting layer (HTL) in p-i-n perovskite solar cells. However, despite its excellent optoelectronic properties, most of the PEDOT:PSS-based $CH_3NH_3PbI_3$ solar cells were reported to have an inferior open-circuit voltage ($V_{OC}$) (0.88-0.97 V)[13] compared to that obtained by n-i-p $CH_3NH_3PbI_3$-based solar cells (1.05-1.12V).[14] Therefore, it is critical to pursue materials and interface systems with minimal voltage losses for developing high-performance inverted perovskite solar cell devices. As alternatives to PEDOT:PSS, other polymeric p-type materials, have been identified as suitable HTLs for high performance p-i-n perovskite solar cells[15] and in most cases lead to devices with higher Voc values. However, the excessive cost of these materials, because of their synthetic procedure and high-purity requirement, is immensely hampering the future commercialization of perovskite solar cells. Low-cost, metal oxides have been identified as promising alternative HTLs in inverted perovskite solar cells.[16,17] Solution processed CuO,[18,9] NiO[19] and CuSCN[20] have been reported as efficient HTLs providing high Voc values (>1 V) for p-i-n perovskite solar cells .

Despite their excellent progress, all the hole transporting layers usually exhibit low electrical conduction properties. By using metal oxide-based HTLs with relatively high electrical conductivity, thicker layers might be used without significant losses in device internal resistances.[21] This is a widely known issue in organic photovoltaic (OPV) technology where the use of thicker charge selective layers are essential for a reliable layer formation using roll-to-roll (R2R) large scale processing as well as device reproducibility.[22] Similar to OPV technology, the use of conductive and thicker charge selective layers results in a series of benefits concerning product development targets of printed perovskite solar cells.[23] In contrast to solution processed n-type metal oxide charge selective layers, the realization of solution processed conductive and transparent p-type metal oxide thin films to be used as HTLs in inverted perovskite solar cells is a challenge. To this end, only few solution processed metal oxide materials which combine high transparency and electrical conductivity have been used as HTLs in p-i-n perovskite solar cells. Recently, Jung *et al.* showed a solution processed CuNiOx,[24] as an efficient HTL for efficient p-i-n perovskite solar cells.

Furthermore, Chen *et al.* demonstrated that the use of a conductive $Ni_xMg_{1-x}O$ HTL leads in an efficient extraction of positive carriers, and the corresponding perovskite-based p–i–n solar cells result in excellent fill factor (FF) values exceeding 80% and hysteresis–free behavior.[25]

Our search for highly conductive and transparent p-type oxide semiconductors for potential use as HTLs in perovskite solar cells, led us to the delafossite compounds, one of the few known categories of p-type materials that combine relatively high electrical conductivity and high transparency in the visible spectrum. Delafossites have the general formula of $A^{+1}X^{+3}O_2$, where A is mostly Cu (or Ag), X is often Al, Ga, Cr and O is oxygen. Delafossites can be crystallized to either a rhombohedral 3R-(*R*3h*m*) or hexagonal 2H-(*P*63/*mmc*) structure.[26] Kawazoe et al. were the first to report that $CuAlO_2$ produced by laser ablation exhibits room temperature electrical conductivity of 1S/cm,[27] one of the highest reported room temperature electrical conductivity values of p-type metal oxide semiconductors. In addition, several studies show energy levels compatible with p-type electrodes of electronic devices, such as large optical band gap[28] and deep work function in the range of 5.2-5.5 eV.[29] However, despite their excellent potential, a major drawback for a wider use of these materials is the extreme amount of energy required to obtain pure delafossite functional films. For example, sol-gel processes can be used to fabricate delafossite thin films only by annealing the gel-films at temperatures over 1000 ºC under inert atmospheres,[30] which is unlikely to be used for the fabrication of solution based perovskite solar cells. Delafossite nano-powders can also be fabricated using hydrothermal methods or microwave assisted hydrothermal methods under extreme pressures (>300 MPa).[31] Wang *et al.* demonstrated that solution-processed $CuGaO_2$ nanoplates is a promising HTL at the top electrode of inverted organic solar cells.[32] However, the large size of $CuGaO_2$ nanoplates result in rough films and expected to affect device reproducibility. To overcome this issue, the same authors demonstrated a suspension of sub-10 nm $CuCrO_2$ NPs, which allows a smooth and transparent film to be fabricated from the dispersion and be used as an efficient HTL for P3HT:PCBM – based organic solar cells.[33]

Despite the impressive development on this promising category of p-type materials, only recently studies show a successful incorporation of delafossite materials as HTL in perovskite solar cells. Zhang *et al.* demonstrated an efficient $CuGaO_2$ nanoplate dispersion fabrication by microwave assisted hydrothermal method as a HTL at the top electrode of n-i-p $CH_3NH_3PbI_3$-based solar cells.[34] In addition, Dunlap-Shohl et al. showed an efficient $CuCrO_2$ HTL in p-i-n perovskite solar cells.[35] We have recently shown that ultrafine delafossite $CuGaO_2$ nanoparticles can be synthesized using a surfactant assisted hydrothermal reaction and be used efficiently as HTLs in p-i-n perovskite solar cells. Specifically, we have presented for the first time that the addition of pluronic P123 surfactant is beneficial for the synthesis of $CuGaO_2$ NPs, as it can decrease the surface energy and temperature of crystallization and control NPs size (~5 nm).[36] DMSO was subsequently used as a ligand and dispersing solvent for stabilizing the $CuGaO_2$ which were successfully used as HTL in p-i-n perovskite solar cells. The corresponding 15 nm $CuGaO_2$-HTLbased inverted perovskite solar cells show PCE of 15.3 %.[36]

$CuAlO_2$ is one of the most promising delafossite materials for optoelectronic applications for several reasons. For example, it exhibits superior optical transparency in the visible spectrum compared to $CuCrO_2$ and at the same time maintains high electrical conductivity.[37] In addition, it is comprised by earth abundant components (Cu and Al) promoting it in a better choice related with large scale use compared with the expensive $CuGaO_2$. To our knowledge there is no reported study on high performance crystalline delafossite $CuAlO_2$ HTLs for solar cell applications. One reason for this, is the complex aluminum chemistry that is prohibiting the efficient hydrothermal synthesis of $CuAlO_2$ nano-powders.[38] Recently Igbari *et al.* used d.c. magnetron sputtering technique 'amorphous' type $CuAlO_2$ (a: $CuAlO_2$) interfacial layer to modify the interface of ITO and PEDOT:PSS in planar perovskite solar cells.[39] Experimental results of crystalline $CuAlO_2$ are limited due to the difficulty to synthesize pure phase Delafossite $CuAlO_2$ thin films.[38] It has been shown that p-type defects play a key role on the electrical properties of $CuAlO_2$[40] and that $CuAlO_2$ thin films can be engineered by nonisovalent Cu-O alloying.[41]

Here we report for the first time the application of a solution processed $CuAlO_2$/Cu-O HTL for p-i-n perovskite solar cells. The XPS peak analysis reported within section 2.1 estimated that the percentage of $CuAlO_2$ is 77% and that of Cu-O is 23% (within a ± 5 % standard error). The $CuAlO_2$/Cu-O powder was developed by high temperature flame spray pyrolysis and the stabilized $CuAlO_2$/Cu-O dispersion sonicated and centrifuged as described in detail below and within the experimental section. Using the supernatant part of the treated $CuAlO_2$/Cu-O dispersions the ultra-large $CuAlO_2$/Cu-O agglomerates are avoided. The proposed treatment is essential to avoid shorting of the devices and to provide functional small area solution processed $CuAlO_2$/Cu-O films. A detail study of material structure, agglomerates/particle sizes, dispersion treatment process, properties, processing and device parameters is presented. The photoluminescence spectroscopy measurements on perovskite films shown a reduced electron-hole pair recombination for the treated $CuAlO_2$/CuO HTL device structure and the electro-impedance spectroscopy studies confirm their efficient hole extraction properties. The experimental results show the potential of hole carrier selectivity and surface properties of the treated $CuAlO_2$/Cu-O HTL providing negligible hysteresis inverted perovskite PVs with best achieved PCE of 16.3%.

## 2. RESULTS AND DISCUSSION

The development of the proposed flame spray pyrolysis solution based HTL includes the following process. For flame spray pyrolysis a metal organic precursor was prepared by a mixture of organic copper and aluminum salt dissolved in a 1:1 mixture of alkyl carboxylic acid and toluene. The metal organic precursors were filtered through a 0.45 mm PTFE syringe filter and fed (5 mL/min) to a spray nozzle, dispersed by oxygen (7 L/min) and ignited by a premixed methane–oxygen fame ($CH_4$: 1.2 L/min, $O_2$: 2.2 L/min). The off-gas was filtered through a glass fiber filter by a vacuum pump at about 20 $m^3$/h. The obtained nano-powder was collected from the glass fiber filter. The flame spray pyrolysis powders undergo a high temperature (1100 °C for 30 min in air) post-treatment sintering process. However, the high temperature sintering step leads to agglomeration and to the formation of large agglomerates (rods/plates) which are shown in supplementary, Figure S1. The powders are stabilized in a 10% wt 2-propoxyethanol (2-PPE) dispersion using an Avantama Ltd surfactant and the stabilized dispersions sonicated for over two hours (with high frequency sonicator probe) following by centrifugation (at 3000 rpm) for 4 min. The supernatant of the treated dispersion was collected and used for the processing of the functional HTLs. By applying high-energy sonication to the dispersions, although large particles are not eliminated other large agglomerates significantly reduced. During the centrifugation process the larger/heavier agglomerates remain at the bottom of the solution. Thus, by making devices from the top of the dispersion (supernatant dispersion) although broad distribution of particle sizes is still existing large agglomerates are in most cases avoided (we discuss this issue in detail later within the text). To form the proposed HTL, the supernatant part of the treated (sonicated and centrifuge) dispersions were dynamically spin coated at 3000 rpm for 60 s on the preheated ITO substrates followed by annealing on a hot plate at 400 °C for 20 min in ambient conditions. Thus, the following terminology is used within the text: Solution A (Room Temperature) named (*Sol A@RT*) is the spray pyrolysis powders stabilized in a 10% wt 2-propoxyethanol using an Avantama Ltd surfactant, Solution B (Room temperature) named (*Sol B@RT*) is the Sol A solution treated (sonicated and centrifuged) at room temperature (RT) and (*Sol B@400ºC*) is Sol A treated (Sonicated and centrifuged) and annealed at 400 °C for 20 min. The term Film A and Film B indicates that the measurements performing in films using the corresponding solutions described above. The term *Film A* (*@RT and @400ºC*) and Film *B* (*@RT and @400ºC*) shows the annealing conditions of the films. We note that (*Film B@400ºC*) represents the identical processing conditions that used on the fabrication step of the HTL within the reported functional inverted perovskite solar cells.

**2.1 CuAlO$_2$ /Cu-O Dispersions and Thin Films**

Figure 1a shows the XRD patterns for the *Sol A@RT*, *Sol B@RT* and *Sol B@400ºC* samples described in detail above, which correspond to the CuAlO$_2$ phase with hexagonal structure (JCPDS 77-2493). These results suggest that the treatment process (sonication and centrifugation) as well as the thermal annealing (up to 400 ºC for 20 min) of sample do not alter its crystallinity. However, it should be stressed that the XRD peaks of polycrystalline CuO are overlapping with the peaks of the CuAlO$_2$ phase and like previous studies, they are not resolved in the XRD patterns of CuAlO$_2$.[40]

Figure 2b shows the region of Cu 2p peak of the XPS spectra of the proposed hole transporting layers. We note that the measurements performed on films (*Film B@400ºC*) using the identical conditions for the processing of the HTLs within the solar cells reported. The Cu 2p$_{3/2}$ and Cu 2p$_{1/3}$ core-level peaks are observed at 932.6 eV and 952.6 eV binding energy respectively, with a spin-orbiting splitting of about 20 eV. Between these two peaks a shape-up satellite exists, which is an indication that some of the Cu atoms are present in the CuO or/and Cu (OH)$_2$ form. The Cu 2p doublet is well fitted by using two peaks, which are assigned to the contributions from the Cu(I) ions (at 932.6 eV binding energy) of the CuAlO$_2$ phase and the Cu(II) ions (at 935.1 eV binding energy) of CuO and Cu(OH)$_2$.[38,42] Based on the XPS peak analysis the estimated percentage of Cu(I) is 77% and that of Cu(II) is 23% (within a ± 5 % standard error).

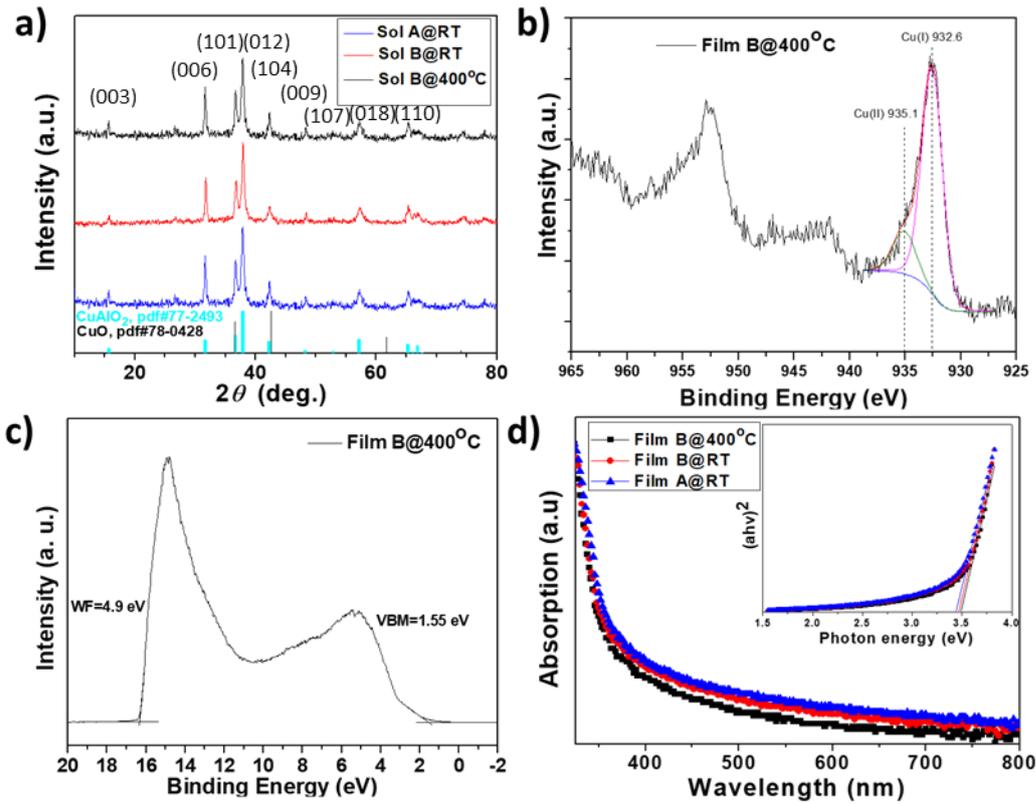

**Figure 1. a)** XRD spectra of the solution A at RT (blue), synthesized by spray pyrolysis and sintered at 1100 ºC, solution B treated at RT (red) and solution B treated and annealed at 400 ºC for 20 min (black) (powders). **b)** XPS spectrum of the Cu2p region for the Film B, fabricated at 400 ºC for 20 min on ITO substrates (films). **c)** Ultra-violent photoelectron spectroscopy (UPS) spectrum of the Film B fabricated at 400 ºC for 20 min on ITO substrates (films). **d)** Optical absorption spectrum of Film A fabricated at RT (blue triangles), Film B fabricated at RT (red circles) and Film B fabricated at 400 ºC for 20 min (black squares), films on quartz substrates. The inset in panel shows the corresponding $(\alpha h\nu)^2$ vs. energy ($h\nu$) plot.

The main part of Cu(II) species is due to the CuO clusters and/or Cu-O dimers of hybridized $CuAlO_2$ lattice, while a minor contribution of $Cu(OH)_2$ phase cannot be excluded. However, due to the absence of CuO crystalline peaks in the XRD patterns of $CuAlO_2$ (Figure 1a), the major part of Cu(II) is attributed to be in the form of Cu-O dimers hybridized in $CuAlO_2$ lattice, which is consistent with previous reported results.[40,41] Therefore, the proposed HTL for inverted perovskite solar cells is termed as $CuAlO_2$/Cu-O.

As shown in Figure 1c, we measured the UPS spectrum of the $CuAlO_2$/Cu-O HTL (*Film B@400°C*). The work function 4.9 eV was calculated by subtracting from the energy of HeI the high binding energy cut-off, which was determined by linear extrapolation to zero of the secondary electrons edge. While the valence band maximum was resolved by linear extrapolation of the leading edge of the UPS spectra in the Fermi region, the ionization potential was calculated to be 6.45 eV. Based on computational results, it was suggested that the valence band region of $CuAlO_2$ is dominated mainly from the Cu 3d orbitals and to some extend from O 2p states.[43,44] Figure 1d shows the absorption spectra on the films which are remain unaffected by the propose treatment and HTL annealing conditions at 400 °C. The band gap energy of the material is estimated in the range of 3.5 eV from the plot of $(ahv)^2$ versus $hv$ (see inset of Figure 1d). The values for ionization potential and band gap estimated here were in the range of values reported for $CuAlO_2$ nonisovalent Cu-O alloying films that have been used for the fabrication of top gate thin film transistors.[41]

The UPS measurements have shown that the ionization potential is calculated at 6.45 eV and the fermi level at 4.9 eV in respect to vacuum. Combining these results with the absorption measurements, where an optical band gap was calculated at ~3.5 eV, we can energetically position the conduction band minimum at ~2.95 eV. The presence of Cu-O dimers clearly affecting the energy bands levels and electronic properties of $CuAlO_2$.[40,41,43,44] Thus, $CuAlO_2$/Cu-O HTL can efficiently block the photogenerated electrons charges at the CB of perovskite layer which are positioned at ~ 3.9 eV in respect to vacuum.[45] On the other hand, hole selection and transport is also influenced by the easily formed acceptor states in the band gap of the $CuAlO_2$ ascribed to Cu(II) and observed in the range of 23% from the XPS measurements shown in Figure 1b and other publications.[40,41] Such states within the bandgap have been observed at around 0.8 eV in respect to VB.[46] The efficient photovoltaic performance results reported in section 2.3 suggest that these states can assist on the selection and transport of the photogenerated holes at perovskite's VB which is located at ~5.4 eV in respect to vacuum.[46]

## 2.2 CuAlO$_2$/Cu-O Dispersion Treatment, Particles Size Distribution and Surface Topography

Figure 2 shows the SEM images on films (*Film A@400°C* and *Film B@400°C*). Figures 2a and 2b show the SEM studies on films fabricated from solution A (*Film A@400°C*). The SEM image of the films fabricated from solution A (Figure 2a with higher magnification ~85.800x) reveals the presence of large agglomerates (dimensions in the range of 850 nm length and diameter 150 nm, as shown in Figure 2a). Despite the large CuAlO$_2$/Cu-O agglomerates, the powders were successfully dispersed in a stable 10% wt 2-PPE solution via the use of an Avantama surfactant (*Sol A@RT*). Figure 2b shows the SEM studies on films fabricated from solution A (*Film A@400°C*) with lower magnification (~12.000x). The very large agglomerates dominated the film topography. A very rough surface, with sharp spikes over 700 nm is observed, which is shown in the profilometry measurements included within Figure S2. As a result, the devices made from the untreated dispersion (*Film A@400°C*) were shorted. A hero device fabricated by using the as prepared flame spray pyrolysis based CuAlO$_2$/Cu-O dispersions resulted in poor device performance as shown in Figure S3.

To achieve functional CuAlO$_2$/Cu-O HTLs the Sol A dispersions were treated with a sonicator probe at high frequencies for over two hours. In total, more than 60 kJ of energy has been delivered to the dispersion, aiming to force some of the large agglomerates to break into smaller particles. Then, the treated dispersion was centrifuged at 3000 rpm for 4 min. By applying high-energy sonication to the dispersions (*Sol B@RT*), although large agglomerates are not eliminated (please see Figure 2c) other agglomerates significantly reduced in size. We believe that during the centrifugation process the larger/heavier agglomerates remain at the bottom of the solution and thus by using the top part of the treated dispersion (supernatant dispersion) to fabricated HTLs large agglomerates are in most cases avoided. Figures 2c and 2d show the SEM surface topography of the treated (according to the above process) $CuAlO_2$/Cu-O thin film annealed at 400 °C for 20 min in air (*Film B@400°C*) with higher (~85.600x) and lower (~12.000x) magnification respectively, showing that large agglomerates sizes have indeed been reduced. The profilometry measurements included within Figure S2 confirm that based on the above process $CuAlO_2$/Cu-O HTL very large sharp spikes can be avoided.

The elimination of the ultra-large agglomerates leads to functional $CuAlO_2$/Cu-O films. Comparing the particle size distribution from the SEM measurements on films (insets of Figure 2b and 2d) we see that indeed the proposed processing steps reduce the size of largest agglomerates from 1 μm (inset of Figure 2b) to 250 nm (inset of Figure 2d).

Using the above proposed treatment (high-energy sonication, centrifugation) and processing steps (using only the supernatant part of the treated CuAlO2/Cu-O dispersion) to avoid the presence of ultra large $CuAlO_2$/Cu-O agglomerates we were able to achieve functional $CuAlO_2$/Cu-O HTLs which are fabricated using supernatant solution processing and temperature annealing at 400 °C for 20 min in air. The particle size distribution from the SEM measurements on films (*Film B@400°C*) fabricated with identical conditions for the preparation of the proposed HTLs (insets of Figure 2d) still shows a large distribution of particle sizes from 25 nm to 250 nm. Thus, even after following the above process 20% of the small area devices reported (active layer of 0.09 mm$^2$) where still shorted. Figure S4 shows a typical device run (active of 0.09 mm$^2$) of total 16 devices. The 3 of the total 16 small area devices where shorted and for this reason have been excluded from the Figure S4. For the functional devices reported within figure S4 the PCE distribution varies from ~11%-14% (average value of 11.5%). However, within the best performing run PCEs in the range of 16 % were achieved (section 2.3). The treated supernatant (CuAlO2/Cu-O) dispersion batch-to-batch reliability and solar cell device performance reproducibility critically depends on the parameters and accuracy of implementation of the delicate CuAlO2/Cu-O functionalization process described within the paper. The proposed treatment results in solution processed CuAlO2/Cu-O HTLs with relatively low electrical resistivity. The electrical conductivity values were measured at $8.5 \times 10^{-2}$ S/cm (Figure S5), which is in good agreement with previous reported measurements of CuAlO$_2$/Cu-O alloy films[40] and larger compared to some of the reported p-type metal oxides used as HTLs in perovskite solar cells.[47]

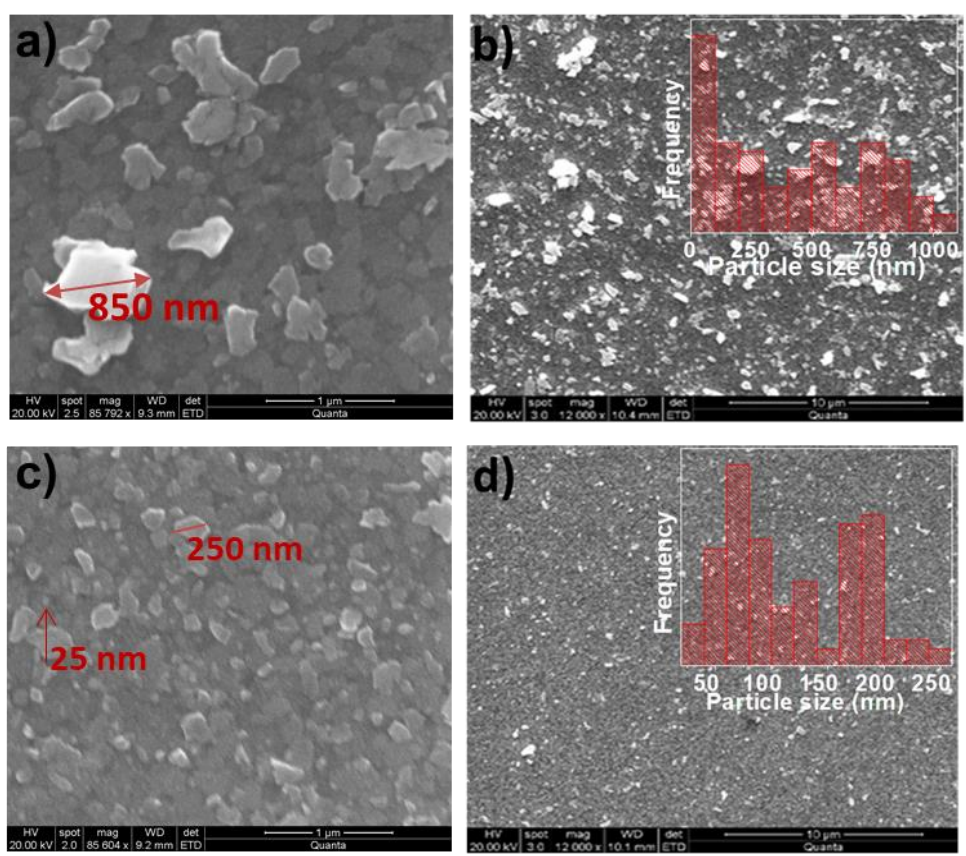

**Figure 2**. SEM images, **a)** top view SEM measurement of CuAlO$_2$/Cu-O thin films (Film A annealed at 400 ºC for 20 min) with ~85.800x magnification and **b)** top view SEM measurement of CuAlO$_2$/Cu-O thin films (Film A annealed at 400 ºC for 20 min) with ~12.000x magnification, inset: size distributions of the CuAlO$_2$/Cu-O particles obtained from corresponding film SEM image. **c)** Top view SEM measurement of CuAlO$_2$/Cu-O thin films (Film B annealed at 400 ºC for 20 min) with ~85.600x magnification and **d)** top view SEM measurement of CuAlO$_2$/Cu-O thin films (Film B annealed at 400 ºC for 20 min) with ~12.000x magnification, inset: size distributions of the CuAlO$_2$/Cu-O particles obtained from corresponding film SEM image.

The above observations are in good agreement with the large area 50x50 μm tapping mode AFM measurements on ITO/[CuAlO$_2$/Cu-O] films showing in Figure 3. Figure 3a show the AFM images of the *Film A@400ºC* and Figure 3b shown the AFM images of *Film B@400ºC* (identical to the optimized HTL). The surface topography of the *Film A@400ºC* fabricated from the untreated CuAlO$_2$/Cu-O dispersion shows RMS=43.7 nm with max peak 727 nm while the surface topography of CuAlO$_2$/Cu-O HTLs (*Film B@400ºC*) show RMS=19.7 nm with max peak 206 nm.

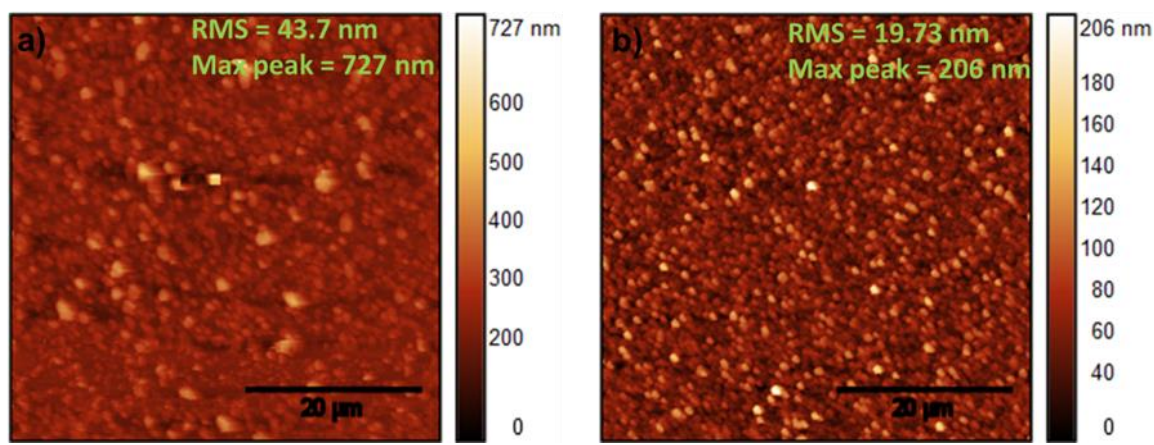

**Figure 3.** Large area 50x50 μm tapping mode AFM measurements of a CuAlO$_2$/Cu-O based thin films **a)** fabricated from solution A (Film A, annealed at 400 ºC for 20 min) on ITO substrates and **b)** fabricated from solution B (Film B annealed at 400 ºC for 20 min) on ITO substrates. The scale bar is 20 μm.

The proposed treatment (high-energy sonication, centrifugation) and processing steps (using only the supernatant part of the treated CuAlO$_2$/Cu-O dispersion) reduced RMS and max peak values. Similar observations are indicated from the profilometer measurements and higher magnification AFM images presented within figure S6 upper plots. The following perovskite active layer step is adequate to cover the treated CuAlO$_2$/Cu-O HTL particle sizes in most cases allowing functional small area inverted perovskite PVs incorporating the treated CuAlO$_2$/Cu-O HTL (Figure S6, bottom plot).

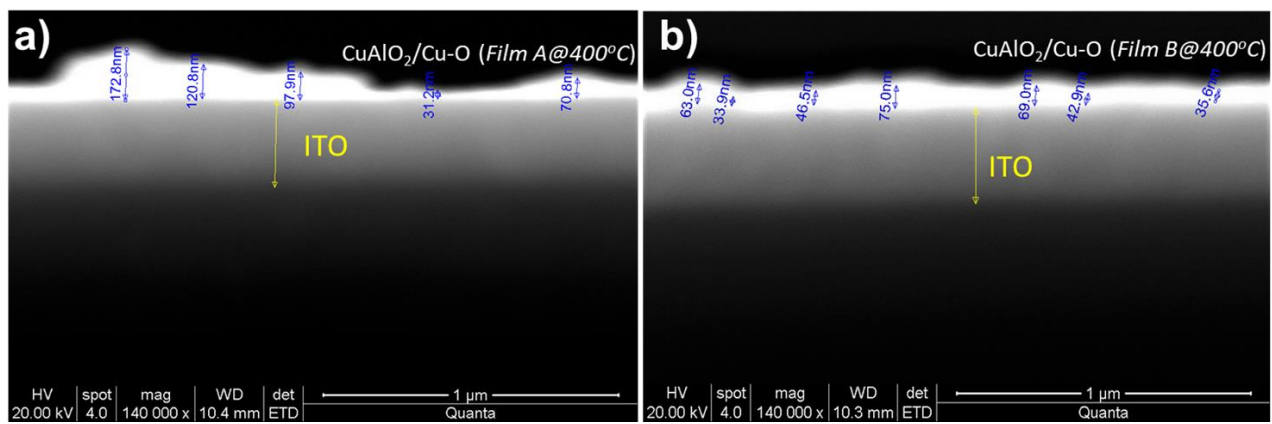

**Figure 4**. Cross section SEM images, **a)** cross section view SEM measurement of CuAlO$_2$/Cu-O thin films fabricated on ITO substrates using solution A (Film A annealed at 400 ºC for 20 min) and **b)** cross section view SEM measurement of CuAlO$_2$/Cu-O thin films fabricated on ITO substrates using solution B (Film B annealed at 400 ºC for 20 min).

Figure 4 illustrates the cross-section SEM images of CuAlO$_2$/Cu-O films on ITO substrates (*Film A@400ºC*) and (*Film B@400ºC*), respectively. Figure 4a shows the cross-section SEM image on CuAlO$_2$/Cu-O film (*Film A@400ºC*), fabricated from solution A (untreated solution), it is shown that due to the presence of large agglomerates, a non-uniform film ranging from 30 nm to 170 nm is formed. As discussed above using such not-treated CuAlO$_2$/Cu-O HTLs (*Film A@400ºC*) resulted to non-functional (shorted) perovskite PVs. On the other hand, Figure 4b shows the cross-section SEM image on CuAlO$_2$/Cu-O film (*Film B@400ºC*), fabricated from solution B treated solution (high-energy sonication, centrifugation and processed using only the supernatant part of the treated CuAlO$_2$/Cu-O dispersion). It can be seen clearly that the proposed treatment of CuAlO$_2$/Cu-O dispersion forms a more uniform film (*Film B@400ºC*) with smaller agglomerates ranging from 30 nm to 70 nm compared to the un-treated film (*Film A@400ºC*). As discussed in detail above and later within the text even though reliability issues are still present using the treated CuAlO$_2$/Cu-O HTLs (*Film B@400ºC*) resulted to functional perovskite PVs when the following perovskite active layer step is adequate to cover the particle distribution of the treated CuAlO$_2$/Cu-O HTL (Figure S6, bottom plot).

## 2.3 CuAlO$_2$ /Cu-O Hole Transporting Layers for Inverted Perovskite Solar Cells

The previous paragraphs describe how we fabricate functional CuAlO$_2$/Cu-O films. To implement those films in solar cell stack, several other requirements must be fulfilled. Figure 5 shows the cross-sectional view SEM image of the treated, CuAlO$_2$/Cu-O HTL (*Film B@400ºC*) perovskite solar cell device structure. The thickness variation due to the broad particle distribution of the treated CuAlO$_2$/Cu-O HTL with estimated average thickness ~50 nm, is indicated within figure 5 and is discussed in more details within the above sections. The perovskite active layer has a thickness of ~260 nm, while the PC$_{70}$BM and the AZO films have a thickness of ~75 and ~50 nm, respectively. The perovskite solar cells have been fabricated on CuAlO$_2$/Cu-O HTL (*Film B@400ºC*) film coated ITO-glass substrates. Finally, 100 nm of Aluminum were thermally evaporated to make an Ohmic contact (not shown in Figure 5).

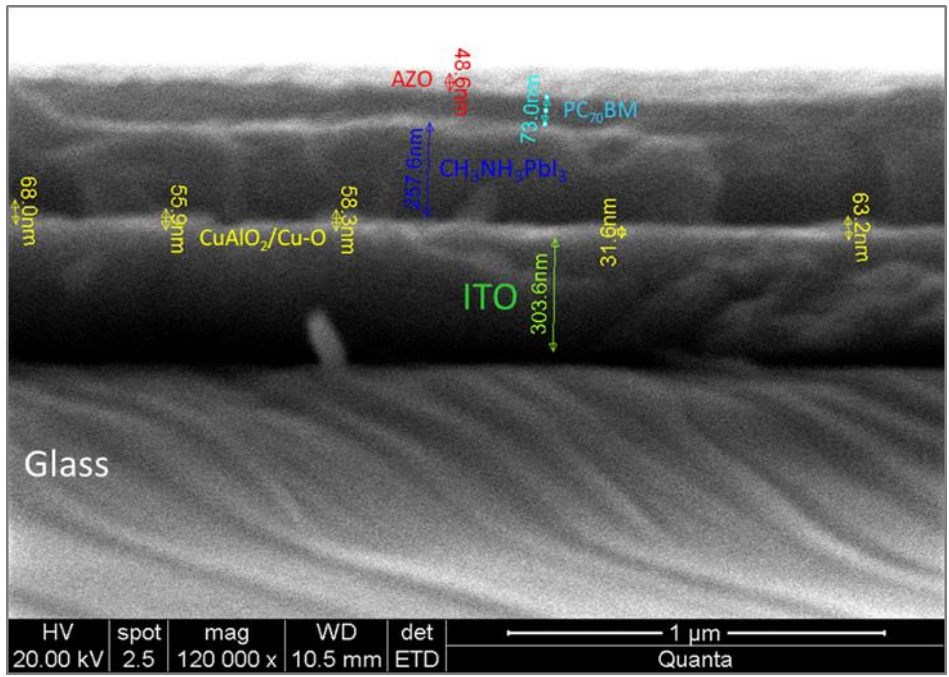

**Figure 5:** Cross section SEM image of the p-i-n device architecture. From bottom to top ITO/[CuAlO$_2$/Cu-O]/CH$_3$NH$_3$PbI$_3$/PC$_{70}$BM/AZO. The thicknesses of the CuAlO$_2$/Cu-O hole transporting layer were accurate measured with a Veeco Dektak 150 profilometer. While an estimation of the other layers can be performed by the cross-sectional SEM image.

The device structure of the inverted perovskite PVs is based on ITO/[CuAlO$_2$/Cu-O]/CH$_3$NH$_3$PbI$_3$/PC$_{70}$BM/AZO/Al) and is schematically shown in Figure 6a. The bottom electrode should be transparent to guarantee that high amount of light reaches the active layer. Furthermore, because the bottom electrode serves as the underlayer for the solution-processed perovskite active layer of the solar cell, its surface properties are of high importance.

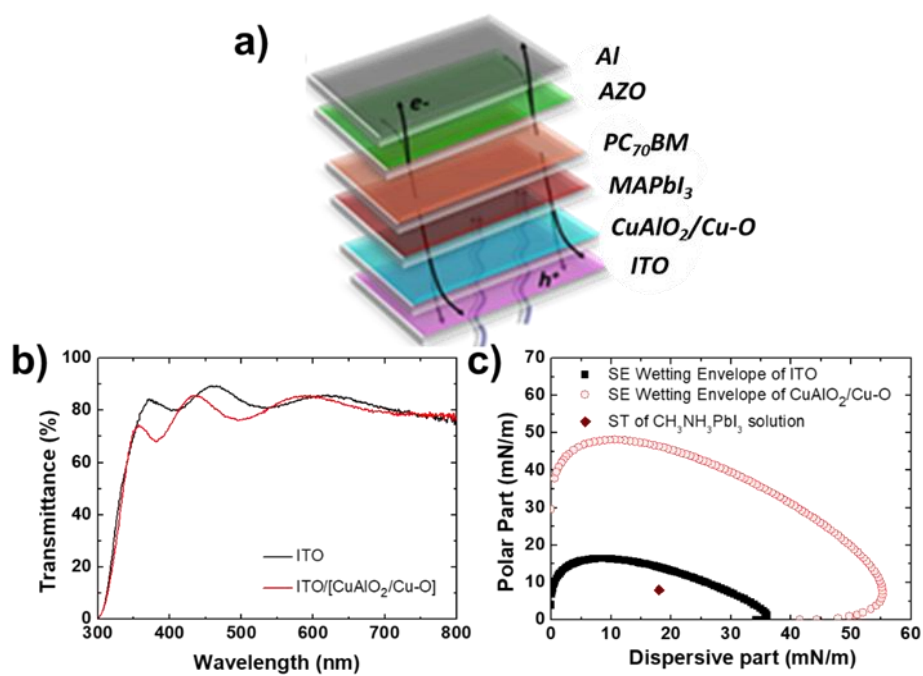

**Figure 6**. **a)** The structure of the p-i-n perovskite solar cells under study (ITO/[CuAlO$_2$/Cu-O]/CH$_3$NH$_3$PbI$_3$/PC$_{70}$BM/AZO/Al). **b)** Transmittance spectra of glass/ITO (black) and glass/ITO/[CuAlO$_2$/Cu-O] (red) and **c)** surface energy wetting envelopes of ITO (black filed squares), ITO/[CuAlO$_2$/Cu-O] (red open circles) and the surface tension of the perovskite solution is also included within the figure.

As shown in Figure 6b, the transparency of the bottom electrode ITO/[CuAlO$_2$/Cu-O] is high, above 80% in the entire visible spectrum (400-800 nm). The high transparency of the proposed ITO/[CuAlO$_2$/Cu-O] bottom electrode guarantees maximum light penetration into the perovskite photo-active layer a parameter which is important for the corresponding photovoltaic devices. Another important requirement for the bottom electrode of perovskite solar cells is its surface wetting properties. The surface free energy wetting envelopes of the two bottom electrodes under study are shown in Figure 6c. Surface energy wetting envelopes are closed curves that are obtained when the polar fraction of a solid is plotted against the disperse part.[48] As shown in Figure 6c the calculated surface tension of the $CH_3NH_3PbI_3$ precursor solution lies well inside the surface energy wetting envelope of the ITO/[$CuAlO_2$/Cu-O] bottom electrode and thus a good wetting is guaranteed. The contact angle of $CH_3NH_3PbI_3$ precursor solution measured at 28º on top of ITO/[$CuAlO_2$/Cu-O] and at 57º on top of ITO, which proves the efficient surface modification of the ITO surface by the $CuAlO_2$/Cu-O HTLs. The digital pictures of the $CH_3NH_3PbI_3$ solution droplets on top the layers ITO and ITO/[$CuAlO_2$/Cu-O] are shown in Figure S7.

In addition, the bottom electrode has a major influence on the perovskite crystallization process as well as the hole collection properties of the p-i-n solar cell. To analyze these important aspects, we performed steady state photoluminescence (PL) on the glass/ITO/$CH_3NH_3PbI_3$ and glass/ITO/[$CuAlO_2$/Cu-O]/$CH_3NH_3PbI_3$ as well as atomic force microscopy (AFM) measurements on glass/ITO/[$CuAlO_2$/Cu-O]/$CH_3NH_3PbI_3$ device structure as shown in Figure 7a and 7b respectively.

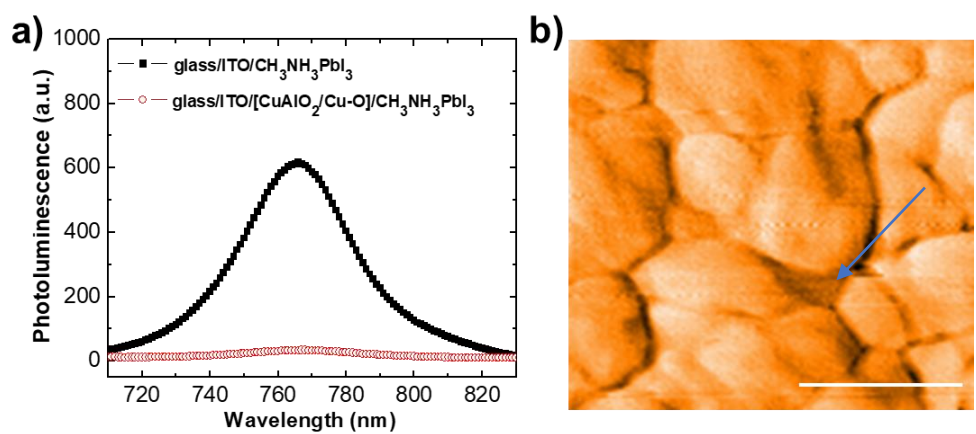

**Figure 7**. **a)** Photoluminescence spectra of Glass/ITO/$CH_3NH_3PbI_3$ and **b)** AFM measurement of $CH_3NH_3PbI_3$ photoactive layers fabricated on top of ITO/[$CuAlO_2$/Cu-O] thin films. The arrow indicates the presence of pin-holes. The scale bar is 500 nm.

As shown in Figure 7a, we compare the steady state PL measurements of $CH_3NH_3PbI_3$ fabricated on top of glass/ITO and on top of glass/ITO/[$CuAlO_2$/Cu-O]. The PL spectra of $CH_3NH_3PbI_3$ fabricated on top of glass/ITO shows a strong peak at 764 nm. The strong PL peak at 764 nm is strongly quenched when the $CH_3NH_3PbI_3$ layer is fabricated on top of ITO/[$CuAlO_2$/Cu-O]. The strong PL quenching of the $CH_3NH_3PbI_3$ PL peak at 764 nm clearly demonstrates efficient charge transfer of the photo-generated species between the $CH_3NH_3PbI_3$ photoactive layer and ITO/[$CuAlO_2$/Cu-O] bottom electrode.

To further investigate the influence of the proposed bottom electrode, we conduct AFM measurements as shown within Figure 7b to study the surface topography of $CH_3NH_3PbI_3$ on top of ITO/[$CuAlO_2$/Cu-O]. The grain size of the polycrystalline $CH_3NH_3PbI_3$ photoactive layers is relatively large with average grain size in the range of 500 nm. The perovskite active layer is found to be relatively smooth with RMS= 11.2 nm (Figure 7b). These observations show that the $CH_3NH_3PbI_3$ photoactive layer is well crystallized. The arrow within figure 7b indicate the presence of perovskite pin-holes most likely due to the broad particle distribution and roughness of the treated $CuAlO_2$/Cu-O underlayer. In agreement with the AFM image of Figure 7b, the corresponding SEM image is also included within the supplementary information (Figure S8). To reduce the negative impact of perovskite active layer pin-holes aluminum-doped zinc oxide (AZO) electron transporting layer have been included within the inverted perovskite PV device architecture. The use of $PC_{70}BM$ and AZO layers within the device architecture assist on reducing the negative effect of pinholes on the inverted perovskite PV device performance.[23]

Based on the above experimental results, we have analyzed the nature of the synthesized particles their broad distribution and discuss the treatment and processing steps needed to fabricate functional $CuAlO_2$/Cu-O HTLs for p-i-n perovskite solar cells. Therefore, p-i-n perovskite photovoltaic devices comprised of ITO/[$CuAlO_2$/Cu-O]/$CH_3NH_3PbI_3$/$PC_{70}BM$/AZO/Al were fabricated using the conditions described in detail above and within the experimental section and analyzed, by using current-voltage (J/V) and external quantum efficiency (EQE) measurements as well as impedance spectroscopy. As noted also above the treated supernatant $CuAlO_2$/Cu-O HTL based solar cell power conversion efficiency reproducibility and reliability critically depends on the parameters and accuracy of implementation of the delicate $CuAlO_2$/Cu-O functionalization process described within the paper. For the device analysis provided below a perovskite PV from the best performing device run is used.

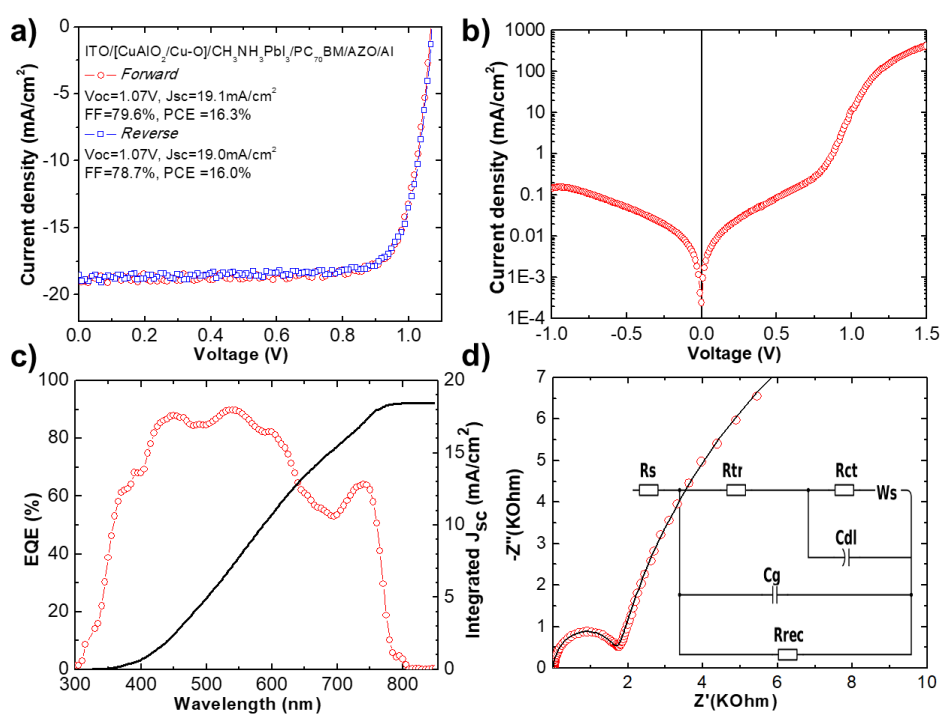

**Figure 8. a)** Current density versus voltage (J/V) characteristics of the best performance PV **a)** under illumination, forward (red open circles) and reverse (blue open squares), **b)** under dark conditions. **c)** External quantum efficiency spectrum for the best performing ITO/[$CuAlO_2$/Cu-O]/$CH_3NH_3PbI_3$/PC [70]BM/AZO/Al solar cell and the integrated photocurrent density of the corresponding perovskite solar cells (right axis). **d)** Nyquist plot for the best performing ITO/[$CuAlO_2$/Cu-O]/$CH_3NH_3PbI_3$/$PC_{70}BM$/AZO/Al solar cell. The insert in Figure 8d shows the equivalent circuit model used to fit the experimental data obtained by impedance spectroscopy.

As shown in Figure 8a the p-i-n device exhibits negligible hysteresis in the J-V plots between the forward and reverse scan directions. The device exhibited Voc=1.07 V, Jsc=19.1

mA/cm$^2$, high FF=79.6% and overall PCE=16.3% in the forward scan with slightly lower overall PCE=16% in the backward scan. The high open circuit Voc value (1.07 V) is similar to other reported metal oxide based HTLs[9,19] and to the value we have achieved using monodispersed CuGaO$_2$ based HTLs for inverted perovskite SCs.[36] We note that in our previous work with solvothermal based synthesized CuO interfacial layers for inverted perovskite solar cells the thickness was in the range of 10 nm due to conductivity and transparency limitations.[9] By the incorporation of the proposed treated CuAlO$_2$/Cu-O HTL we have achieved rougher, non-uniform but thicker (in the range of 50 nm) HTLs. When the following perovskite active layer step is effectively covers the broad particle size distribution of the treated CuAlO2/Cu-O HTL the corresponding perovskite PVs exhibits FF values up to ~80%. This is due to the adequate electrical conductivity of CuAlO$_2$/Cu-O HTL as well as the large grain sizes of CH$_3$NH$_3$PbI$_3$ active layer. As a result, the parasitic resistances, known to limit the FF values of perovskite solar cells (i.e. interface and grain boundaries recombination) is suppressed. The latter is also verified with the measurements of the solar cell in dark conditions (Figure 8b), which shows very good diode behavior both in high negative voltages (parallel resistance regime) and in high positive voltages (series resistance regime). In addition, the solar cell performance show photo-stable performance at maximum power point as shown in Figure S9. Initial lifetime testing under the basic ISOS-D-1 protocol (RT, ambient humidity in the dark) is also performed for the encapsulated inverted PVs showing stable performance for the first 750 hours (Figure S10). Despite the observation that devices show stable performance up to 750 hours under the basic protocol ISOS-D-1 after this time most of the devices were not able to be measured due to Al contact failure (Al oxidation due to environmental conditions). Thus, the above conditions cannot be used to provide a clear understanding on the degradation mechanisms related to bottom electrode effects of the inverted PV solar cells reported within this paper. More detailed accelerated lifetime testing under light soaking will be presented in future publications. We have recently shown that under temperature accelerated lifetime conditions the interaction of the perovskite active layer with the top Al metal electrode through diffusion mechanisms cause a major thermal degradation pathway for inverted perovskite photovoltaics (PVs). We have proposed thick fullerenes diffusion blocking layers to improve the thermal stability of inverted perovskite PVs.[49]

We note that the treated dispersion supernatant batch-to-batch reproducibility critically depends on the sonication and centrifugation parameters used as well as the accuracy of using the top part of the solution after centrifugation process (supernatant dispersion). All the high-performance devices using the $CuAlO_2$/Cu-O HTL reported in this paper achieved by following the implementation of this delicate treatment. We further analyze the device operation by measuring the external quantum efficiency (EQE) as a function of wavelength as shown in Figure 8c. The integrated Jsc (18.5 mA/cm$^2$) value extracted from the EQE spectra is in accordance with the J-V measured Jsc (19.1 mA/cm$^2$) with only 3% deviation is shown within Figure 6c. In addition, the EQE spectra is above 85% in the range 450-550 nm. In this area, the optical absorption of $CH_3NH_3PbI_3$ photoactive layer is almost 100%. Thus, the high EQE in this optical spectrum is a strong indication that almost all the absorbed photons are converted into charges and collected efficiently at the device electrodes. In the region of 550 nm-700 nm the EQE percentage drops from 75% down to 55% which is most likely due to the drop-in absorption strength of the corresponding $CH_3NH_3PbI_3$ photo-active layers. The peak of the EQE spectra at 700-760 nm can be attributed to the use of AZO nanoparticulate layer at the top electrode resulting in light scattering and increased back electrode reflectivity.[23]

To better understand the photo-physical processes and the mechanisms of charge extraction, we conduct impedance spectroscopy measurements for the ITO/[$CuAlO_2$/Cu-O] /$CH_3NH_3PbI_3$/$PC_{70}BM$/AZO/Al solar cells. The Nyquist plots were constructed, and the experimental data were fitted into an equivalent circuit model (ECM) as shown in Figure 8d. The ECM used as well as the physical meaning of its components can be found in more details elsewhere.[50,11] The high frequency response is generally attributed to the charge transport resistance in the carrier selective contacts, whereas the low frequency arc is attributed to the charge recombination resistance of the device. Interestingly, the recombination resistance ($R_{rec}$) value is found to be at ~35000 Ohm, which is considerably higher compared to the values reported for different HTLs using similar ECMs (e.g. PEDOT: PSS),[11] indicating that there is a considerable decrease to the charge recombination events throughout the device interfaces as well as the individual layers themselves. In addition, the double layer capacitance ($C_{dl}$) is relatively small compared to other commonly used HTLs,[11] indicative of reduced charge accumulation at the interfaces and thus faster hole extraction.

The above mechanisms are the major reasons for the large Voc, and FF values measured from the J-V characteristics of the best performing inverted p-i-n perovskite solar cells under study (Figure 8a). Furthermore, in order to obtain good fits, we note that a constant phase element (CPE) was used for $C_{dl}$, which models the behavior of a non-ideal capacitor.[51] CPEs are often necessary to be used to compensate for variations in capacitance arising from phenomena such as non-uniform current distribution and variations in a layer thickness.[52] In our case we believe that it is a case of non-uniform space charge distribution at the metal-oxide based HTL electrode interfaces.[11]

To summarize, the flame spray pyrolysis synthesized $CuAlO_2$/Cu-O HTL (Film A annealed at 400 °C for 20 min) shows very large agglomerates, extremely rough surface, with ultra-sharp spikes. As a result, devices made from the untreated CuAlO2/Cu-O film were not functional (shorted). To achieve functional CuAlO2/Cu-O HTLs a delicate treatment process applied. The proposed CuAlO2/Cu-O dispersion treatment (high-energy sonication, centrifugation) and processing steps (using only the supernatant part of the treated CuAlO2/Cu-O dispersion) of the high temperature flame spray pyrolysis CuAlO2/Cu-O (Film B annealed at 400 °C for 20 min) provided functional HTLs for the small active area (0.09 mm$^2$) perovskite PVs reported within this paper. Even though device reproducibility and reliability issues are still present due to delicate functionalization process applied, the CuAlO$_2$/Cu-O HTL has adequate wetting properties relevant to CH$_3$NH$_3$PbI$_3$ surface tension requirements and forms a functional ITO/[CuAlO$_2$/Cu-O] bottom electrode as confirmed by the efficient quenching of the photoluminescence. The analysis of the impedance spectroscopy measurements indicates reduced interfacial charge recombination and faster charge collection properties. The optimized photoactive layer properties when the CH$_3$NH$_3$PbI$_3$ fabricated on top of ITO/ Treated [CuAlO$_2$/Cu-O] HTL shown a Jsc value of 19.1 mA/cm$^2$ while provided a high Voc of 1.07 V. The above combined with the FF of 79.6% resulted to champion PCE (forward scan) values of 16.3% and negligible hysteresis effect.

## 3. CONCLUSION

We presented a solution processed CuAlO$_2$/Cu-O HTLs and discussed their functional operation in p-i-n perovskite solar cells. The CuAlO$_2$/Cu-O powders are obtained by high temperature flame spray pyrolysis. However, we observe that the large agglomerate sizes of the as prepared CuAlO$_2$/Cu-O dispersions are limiting the application of the corresponding HTL for perovskite solar cell applications (corresponding perovskite PVs were shorted). To achieve CuAlO$_2$/Cu-O HTL functionality, the initial CuAlO$_2$/Cu-O dispersion is treated with high frequency sonicator probe, which force some of the large agglomerates to break into smaller CuAlO$_2$/Cu-O particles. Most of the remaining large particles are avoided from the dispersions by centrifugation and taking the top part of the solution (supernatant dispersion) for fabrication of the corresponding HTLs for the development of functional perovskite PVs.

Despite the broad particle sizes distribution by following the above treatment, we have presented the fabrication of transparent and electrically conductive CuAlO$_2$/Cu-O electronic films. We have shown that by the proposed dispersion treatments and processing steps for the fabrication of the CuAlO$_2$/Cu-O HTL functional small area (0.09 mm$^2$) inverted perovskite–based solar cell can be achieved with PCE up to 16.3%. Identification of synthetic/chemical methods to reduce $CuAlO_2$/Cu-O particle sizes and control the particle size distribution will be needed to further improve the device performance and reliability of inverted perovskite solar cells.

## 4. EXPERIMENTAL SECTION

**4.1. Materials.** Pre-patterned glass-ITO substrates (sheet resistance 4 Ω/sq) were purchased from Psiotec Ltd, $PbI_2$ from Alfa Aesar, methylammonium iodide from Dyesol Ltd, Aluminum-doped zinc oxide (AZO) ink from Nanograde (product no N-20X), PC [70]BM from Solenne BV. All the other chemicals used in this study were purchased from sigma Aldrich.

**4.2. Preparation.** *Dispersions.* The $CuAlO_2$/Cu-O powders were synthesized from Avantama Ltd by flame spray pyrolysis. For flame spray pyrolysis a metal organic precursor was prepared by a mixture of organic copper and aluminum salt dissolved in a 1:1 mixture of alkyl carboxylic acid and toluene. The metal organic precursors were filtered through a 0.45 mm PTFE syringe filter and fed (5 mL/min) to a spray nozzle, dispersed by oxygen (7 L/min) and ignited by a premixed methane–oxygen fame ($CH_4$: 1.2 L/min, $O_2$: 2.2 L/min). The off-gas was filtered through a glass fiber filter (Schleicher & Schuell) by a vacuum pump (Busch, Seco SV1040CV) at about 20 $m^3$/h. The obtained nano-powder was collected from the glass fiber filter. The as-prepared particles were annealed at 1100 ºC for 30 min in air. The powders were stabilized in 2-PPE at 10 wt% using an undisclosed Avantama Ltd surfactant. The treatment of the $CuAlO_2$/Cu-O dispersion by a sonicator probe and the following sonication step was performed at Cyprus University of Technology (CUT). More specifically, to reproduce the manuscript results the flame spray pyrolysis based $CuAlO_2$/Cu-O dispersions must be treated with high frequency sonicator probe for over two hours and centrifuged at 3000 rpm for 4 min. Importantly after the centrifuged step only the supernatant dispersion should be used in order to achieve functional devices as reported here.

*Device Fabrication.* The p-i-n solar cells under study had the structure ITO/[CuAlO$_2$/Cu-O]/CH$_3$NH$_3$PbI$_3$/PC$_{70}$BM/AZO/Al. ITO substrates were sonicated in acetone and subsequently in isopropanol for 10 min and heated at 120 ºC on a hot plate 10 min before use. To form a CuAlO$_2$/Cu-O hole transporting layer, the CuAlO$_2$/Cu-O dispersions used in this study were dynamically spin coated at 3000 rpm for 60 s on the preheated ITO substrates followed by an annealing step on a hot plate at 400 ºC for 20 min in ambient conditions. The perovskite precursor solution was prepared 1 h prior spin coating by mixing lead acetate: methylamonium iodide (1:3) at 36 wt % in DMF with the addition of 1.5 mole % (to PbAc) of methylamonium bromide. The precursor was filtered with 0.1 μm PTFE filters and deposited by static spin coating at 4000 rpm for 60 s followed by annealing for 5 min at 80 ºC. The PC$_{70}$BM solution, 20 mg/ml in chlorobenzene, was dynamically spin coated on the perovskite layer at 1000 rpm for 30 s. AZO ink was dynamically spin coated on top of PC$_{70}$BM at 3000 rpm for 30 s. Finally, 100 nm Al layers were thermally evaporated through a shadow mask to finalize the devices with active of 0.09 mm$^2$. While an estimation of the other layer thicknesses can be performed by the cross-sectional SEM image included in Figure 5. The devices were encapsulated directly after the evaporation of the metal contact, in the glove box using a glass coverslip and an encapsulation epoxy resin (Ossila E131) activated by 365 nm UV-irradiation.

**4.3. Characterization.** The thicknesses and the surface profile of the active layers were measured with a Veeco Dektak 150 profilometer. XRD patterns were collected on a PANanalytical X'pert Pro MPD powder diffractometer (40 kV, 45 mA) using Cu Kα radiation ($\lambda$ = 1.5418 Å). The current density-voltage (J/V) characteristics were characterized with a Botest LIV Functionality Test System. Both forward (short circuit -> open circuit) and reverse (open circuit -> short circuit) scans were measured with 10 mV voltage steps and 40 ms of delay time. For illumination, a calibrated Newport Solar simulator equipped with a Xe lamp was used, providing an AM1.5G spectrum at 100 mW/cm$^2$ as measured by a certified oriel 91150V calibration cell. A custom-made shadow mask was attached to each device prior to measurements to accurately define the corresponding device area. EQE measurements were performed by Newport System, Model 70356_70316NS. Impedance spectroscopy was performed using an Autolab PGSTAT 302N equipped with FRA32M module. To extract the Nyquist plots, the devices were illuminated using a red LED at 625 nm and 100 mw/cm$^2$. A small AC perturbation voltage of 10 mV was applied, and the current output was measured using a frequency range of 1 MHz to 1 Hz. The steady-state DC bias was kept at 0 V. Transmittance and absorption measurements were performed with a Schimadzu UV-2700 UV–vis spectrophotometer. Steady-state photoluminescence experiments were performed in a Fluorolog-3 Horiba Jobin Yvon spectrometer based on an iHR320 monochromator equipped with a visible photomultiplier tube (Horiba TBX-04 module). The PL was non-resonantly excited at 400 nm by the line of 5 mW Oxxius laser diode. Atomic force microscopy (AFM) images were obtained using a Nanosurf easy scan 2 controller in tapping mode. Scanning Electron Microscope (SEM) measurements were performed using a Quanta 200 microscope (FEI, Hillsboro, Oregon, USA). X-ray photoelectron spectra (XPS) and Ultraviolet Photoelectron Spectra (UPS) were recorded by using a Leybold EA-11 electron analyzer operating in constant energy mode at a pass energy of 100 eV and at a constant retard ratio of 4 eV for XPS and UPS, respectively. The spectrometer energy scale was calibrated by the Au $4f_{7/2}$ core level binding energy, BE, (84.0 ± 0.1 eV) and the energy scale of the UPS measurements was referenced to the Fermi level position of Au at a binding energy of 0 eV. All binding energies were referred to the C 1s peak at 284.8 eV of surface adventitious carbon. The X-ray source for all measurements was a non-monochromatized Al Kα line at 1486.6 eV (12 keV with 20 mA anode current). For UPS measurements, the HeI (21.22 eV) excitation line was used. A negative bias of 12.22 V was applied to the samples during UPS measurements in order to separate secondary electrons originating from the sample and the spectrometer. The sample work function was determined by subtracting the high binding energy cut-off from the HeI excitation energy (21.22 eV). The position of the high-energy cut-off was determined by the intersection of a linear fit of the high binding portion of the spectrum with the background. Similarly, the valence band maximum is determined with respect to the Fermi level, from the linear extrapolation of the valence band edge to the background.

## ASSOCIATED CONTENT

**Supporting Information**: Supplementary information includes additional details for materials characterization, surface properties, processing and device performance (Efficiency and initial lifetime studies).

## AUTHOR INFORMATION

**Corresponding Author**

*E-mail: stelios.choulis@cut.ac.cy

## ACKNOWLEDGEMENTS

This project has received funding from the European Research Council (ERC) under the European Union's Horizon 2020 research and innovation programme (grant agreement No 647311).

# Table of Contents Graphic

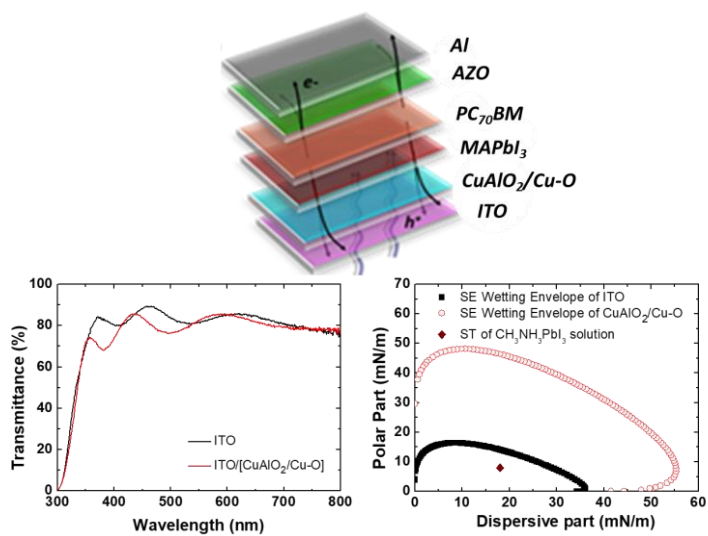

# Supporting Information

## Inverted Perovskite Photovoltaics using Flame Spray Pyrolysis Solution based CuAlO$_2$/Cu-O Hole Selective Contact

By Achilleas Savva,[1] Ioannis T. Papadas,[1] Dimitris Tsikritzis,[1] Apostolos Ioakeimidis,[1] Fedros Galatopoulos,[1] Konstantinos Kapnisis,[1] Roland Fuhrer,[2] Benjamin Hartmeier,[2] Marek F. Oszajca,[2] Norman A. Luechinger,[2] Stella Kennou,[3] Gerasimos Armatas[4] and Stelios A. Choulis*,[1]

[1] Dr A. Savva, Dr I.T. Papadas, Dr Dimitris Tsikritzis, Mr Apostolos Ioakeimidis, Mr Fedros Galatopoulos, Dr Konstantinos Kapnisis, Prof S. A. Choulis,
Molecular Electronics and Photonics Research Unit, Department of Mechanical Engineering and Materials Science and Engineering, Cyprus University of Technology, Limassol, 3603 (Cyprus).

[2] Roland Fuhrer, Benjamin Hartmeier, Marek Oszajca, Dr Norman Luechinger
Avantama Ltd, Staefa, Laubisrutistr. 50, CH-8712, Switzerland

[3] Department of Chemical Engineering, University of Patras, 26504, Patras, Greece

[4] Department of Materials Science and Technology, University of Crete, Heraklion 71003, Greece

*Corresponding Author: Prof. Stelios A. Choulis

E-mail: stelios.choulis@cut.ac.cy

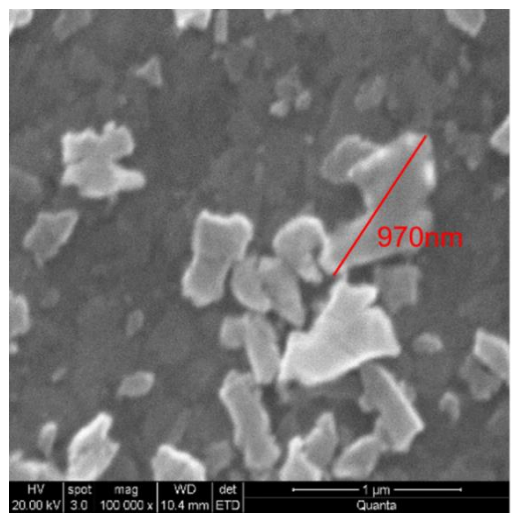

**Figure S1:** Top view SEM measurements of CuAlO$_2$/Cu-O thin films fabricated using solution A and annealed at 400 °C for 20 min [As prepared CuAlO$_2$/Cu-O dispersion (not treated with probe sonicator and centrifuged)].

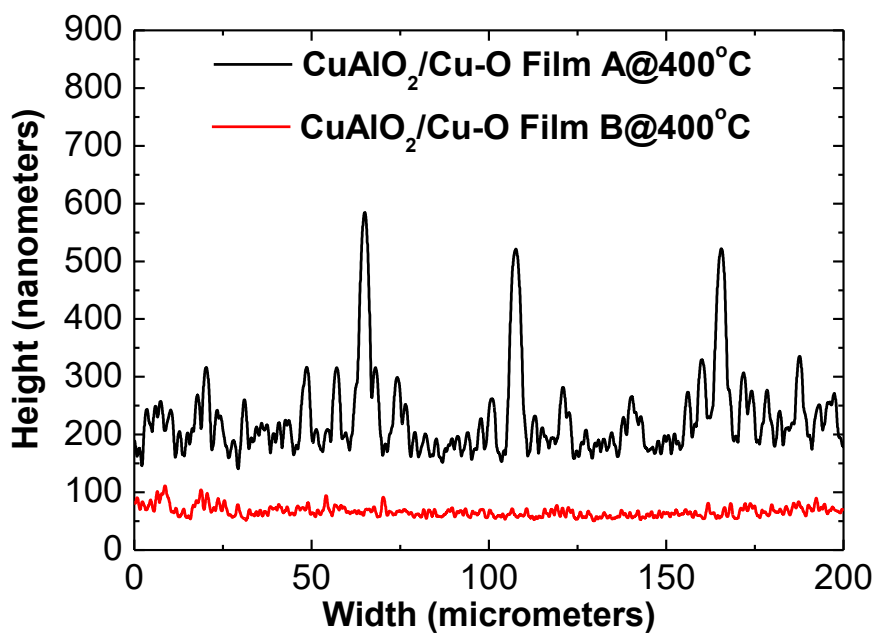

**Figure S2:** Surface profile measurements of CuAlO$_2$/Cu-O thin films fabricated from CuAlO$_2$/Cu-O sol A (black) and CuAlO$_2$/Cu-O sol B (red), annealed at 400 °C for 20 min. As indicated within the main manuscript CuAlO$_2$/Cu-O thin films fabricated from a stable 10

wt% $CuAlO_2$/Cu-O dispersion in 2-PPE ($CuAlO_2$/Cu-O sol A) and the same $CuAlO_2$/Cu-O dispersion treated with high frequency sonicator probed for over two hours, centrifuged at 3000 rpm for 4 min and used the supernatant part of the treated dispersions ($CuAlO_2$/Cu-O sol B).

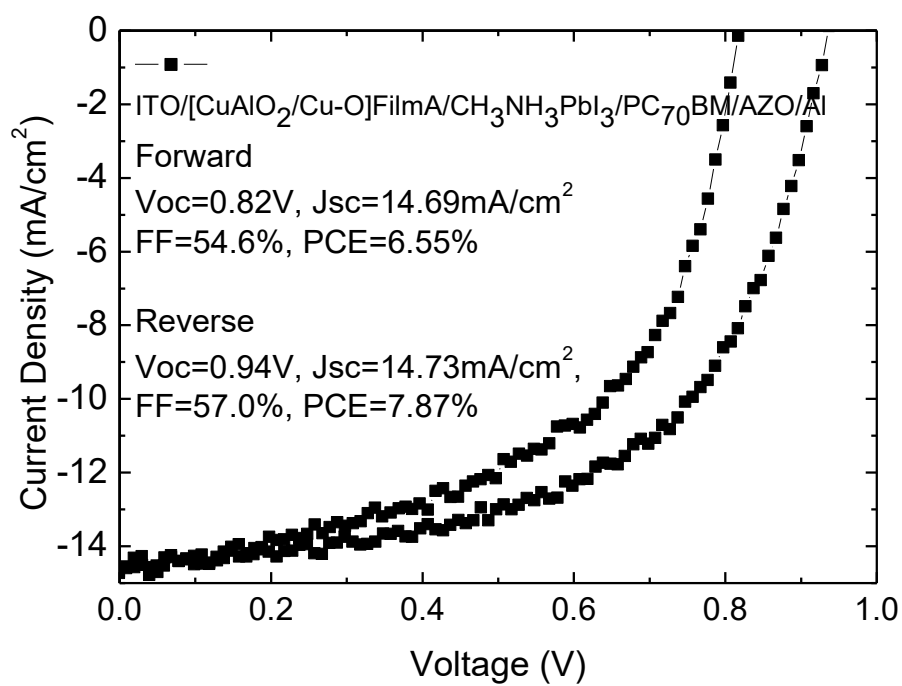

**Figure S3:** J/V characteristics for a hero device, in forward and reverse scan fabricated by using the as prepared $CuAlO_2$/Cu-O solution A (not treated with probe sonicator and centrifuged, annealed at 400 °C for 20 min).

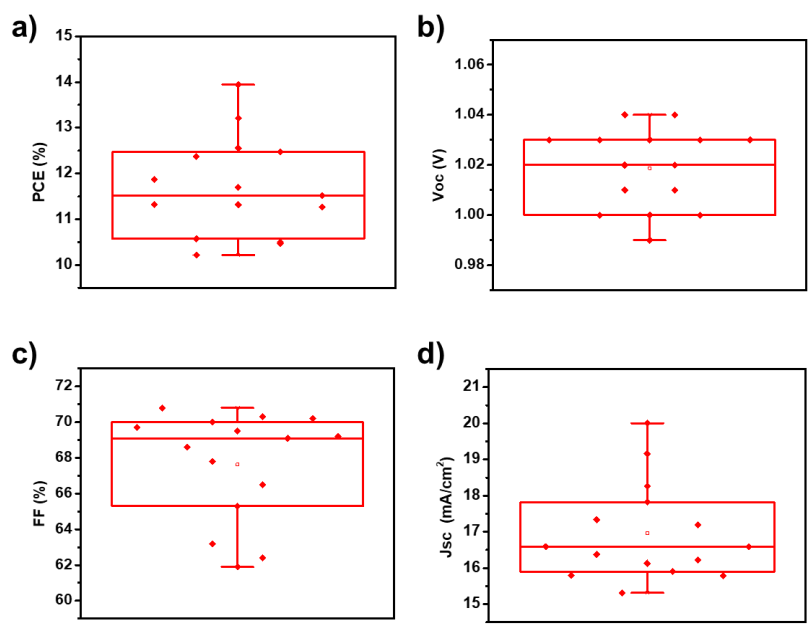

**Figure S4.** Is important to highlight that before performing all the additional characterization studies required during the referee process, we were always made devices to be confident that the measurements performed on functional $CuAlO_2$/Cu-O material/films. An example of a typical photovoltaic run performed during the referee process period is presented on this plot. 3 out of total 16 small area inverted perovskite PVs using [$CuAlO_2$/Cu-O]-HTL cells from the optimized treatment reported within the paper were shorted and thus do not included in the plot. The photovoltaics parameters represented in box plots **a)** power conversion efficiency (PCE), **b)** open circuit voltage (Voc), **c)** fill factor (FF) and **d)** current density (Jsc) are obtained from the total 13 functional devices. We note that treated supernatant ($CuAlO_2$/Cu-O) dispersion batch-to-batch reliability and solar cell device performance reproducibility critically depends on the parameters and accuracy of implementation of the delicate functionalization process described within the paper.

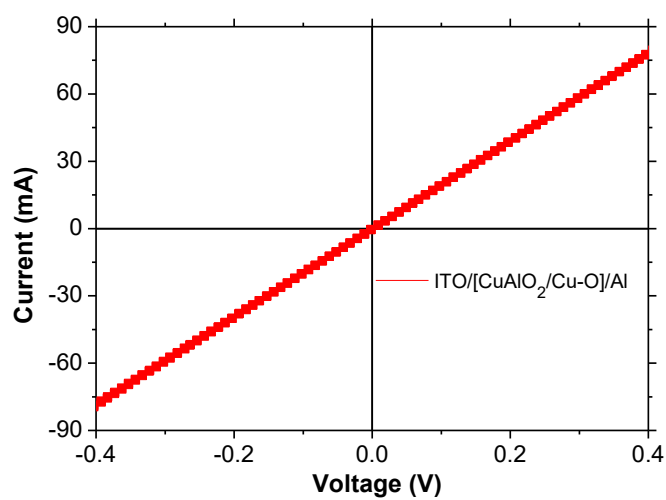

**Figure S5:** I/V curve of $CuAlO_2$/Cu-O *Film B@400ºC* (HTL) sandwiched between 2 contacts (ITO and Al). The slope was used to calculate the resistivity of the thin films under study.

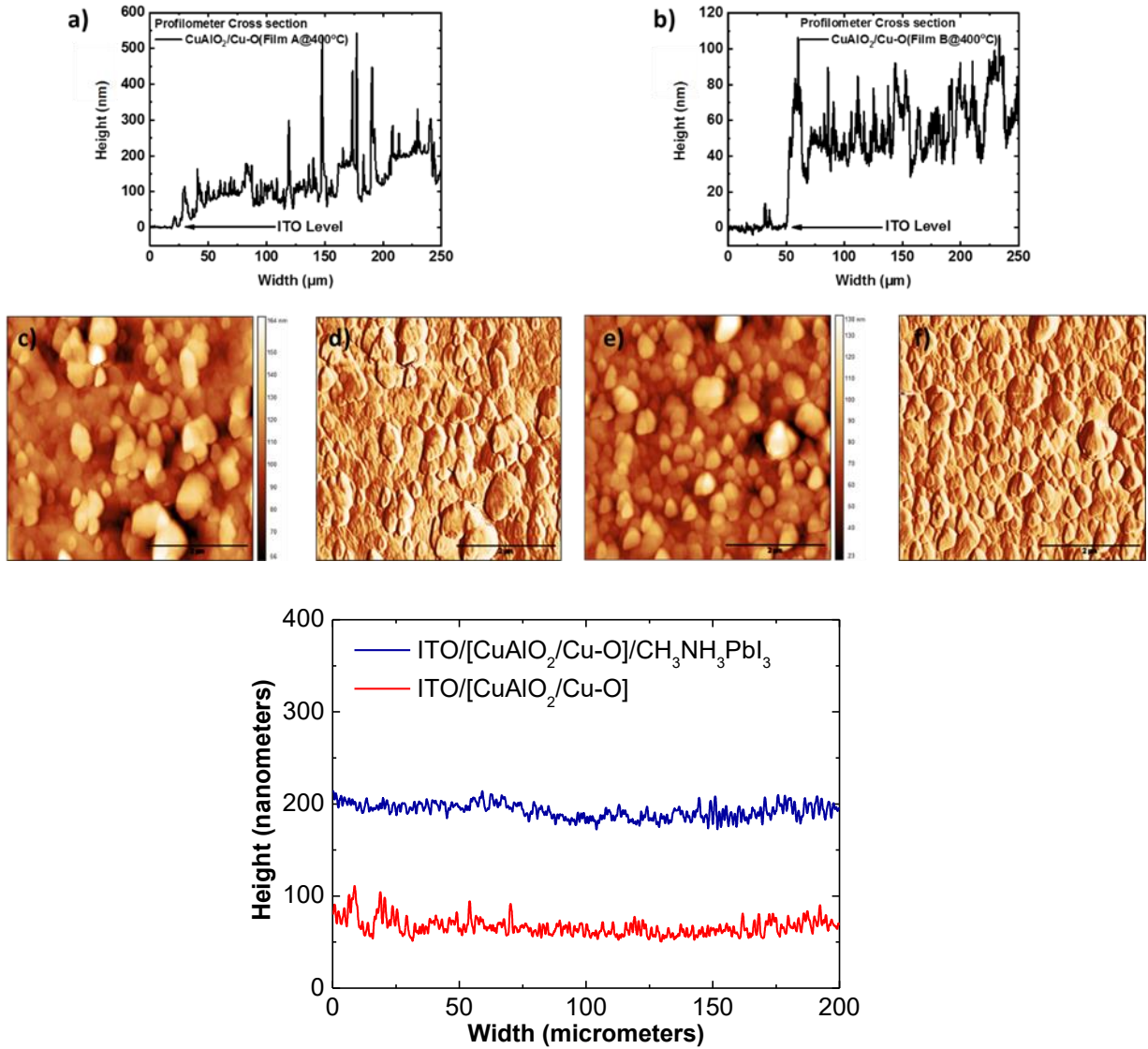

**Figure S6:** *Upper plots (a-f):* Figure a) and b) illustrates the profilometer image of CuAlO2/Cu-O films (Film A and B @ 400°C) fabricated on ITO substrates using solution A (not treated) and B (treated), respectively. In both cases part of the film was mechanically etched and the profile of the films from ITO to ITO/CuAlO2/Cu-O for a range of 250 μm was measured (the line close to zero height at the beginning of the graph corresponds the ITO level). Using the film B @ 400°C proposed treatment (high-energy sonication, centrifugation) and processing steps (using only the supernatant part of the treated CuAlO2/Cu-O dispersion) large agglomerates/spikes are indeed reduced. The AFM measurement (50x50μm) has been presented within the main text (figure 3). Within this figure(figure S6) a larger magnification AFM images (5x5 μm) are shown for each film. Images c, e are the topographies and d,f the corresponding phase images of the Film A 400 °C and Film B@400 °C respectively. The phase images show a narrow phase shift in both cases. ***Bottom Plot:*** Surface profiles of the device structures using the proposed treatment (high-energy sonication, centrifugation) and processing steps (using only the supernatant part of the treated CuAlO2/Cu-O dispersion) of the CuAlO2/Cu-O (Film B annealed at 400 °C for 20 min) HTL. The two device structures are the ITO/CuAlO2/Cu-O and ITO/CuAlO2/Cu-O/CH3NH3PbI3. Despite the rough profile, of the CuAlO2/Cu-O it is shown that the following perovskite active layer shows relative smooth profile. The perovskite processing step is adequate to cover the broad distribution of particle

sizes in most cases for the film B @ 400°C conditions allowing functional small area inverted perovskite PVs incorporating the treated CuAlO2/Cu-O HTL.

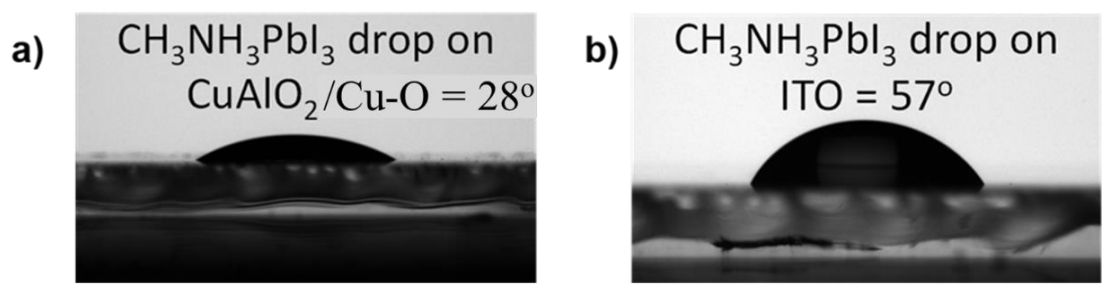

**Figure S7:** Digital photographs of a $CH_3NH_3PbI_3$ solution drop on top of **a)** ITO/[CuAlO$_2$/Cu-O] solution B and **b)** ITO corresponding layers.

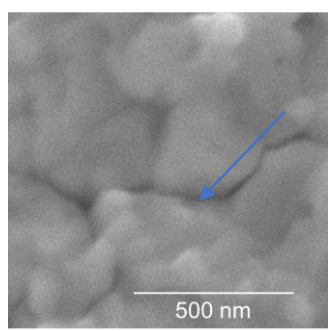

**Figure S8.** Top view SEM image of $CH_3NH_3PbI_3$ photoactive layers fabricated on top of ITO/[CuAlO$_2$-Cu-O] films.

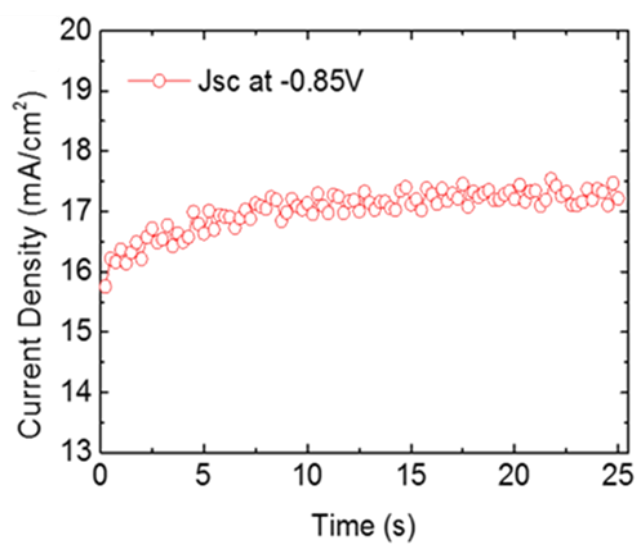

**Figure S9:** Photocurrent stability at the maximum power point of the best performing PV CuAlO$_2$/Cu-O-HTL inverted perovskite based solar cells showing within the main paper text.

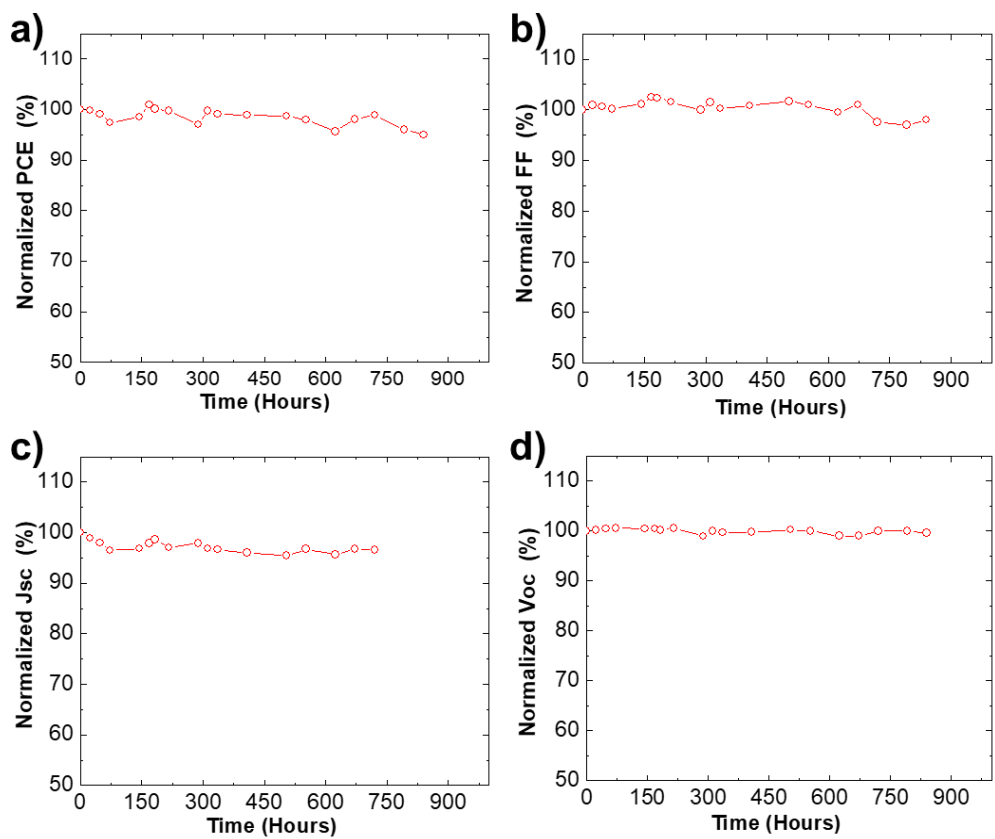

**Figure S10:** Initial lifetime testing of average values of 12 CuAlO$_2$/Cu-O-HTL based inverted perovskite PVs. The plot shows normalized values of **a)** power conversion efficiency (PCE), **b)** fill factor (FF), **c)** current density (Jsc) and **d)** open circuit voltage (Voc), as a function of time of exposure under the basic protocol ISOS-D-1 (RT, ambient humidity in the dark) for the encapsulated p-i-n perovskite-based cells using CuAlO$_2$/Cu-O HTLs (red open circles). Despite the observation that devices show stable performance up to 750 hours under the basic protocol ISOS-D-1 after this time most of the devices were not able to be measured due to Al contact failure (Al oxidation due to environmental conditions). Thus, the above conditions cannot be used to provide a clear understanding on the degradation mechanisms related to bottom electrode effects of the inverted PV solar cells reported within this paper and for this reason lifetime performance under ISOS-D-1 have been included as initial lifetime testing within the manuscript supplementary information.